\documentstyle[12pt,titlepage,epsf]{article}
\begin{document}
\pagestyle{empty}
\baselineskip=0.212in

\begin{flushleft}
\large
{SAGA-HE-95-96  
   \hfill January 31, 1996}  \\
\end{flushleft}
 
\vspace{2.5cm}
 
\begin{center}
 
\LARGE{{\bf Shadowing Aspects}} \\
\vspace{0.3cm}

\LARGE{{\bf of Nuclear Parton Distributions}} \\
 
\vspace{1.5cm}
 
\Large
{S. Kumano $^*$ }         \\
 
\vspace{0.8cm}
  
\Large
{Department of Physics}         \\
 
\vspace{0.1cm}
 
\Large
{Saga University}      \\
 
\vspace{0.1cm}

\Large
{Saga 840, Japan} \\

\vspace{2.0cm}
 
\large
{Talk given at the Joint Japan-Australia Workshop on} \\

\vspace{0.3cm}

{``Quarks, Hadrons, and Nuclei"} \\

\vspace{0.7cm}

{Adelaide, Australia, Nov. 15 -- 24, 1995 (talk on Nov. 20, 1995)}  \\
 
\end{center}
 
\vspace{1.5cm}

\vfill
 
\noindent
{\rule{6.cm}{0.1mm}} \\
 
\vspace{-0.2cm}
\normalsize
\noindent
{* Email: kumanos@cc.saga-u.ac.jp. 
   Information on his research is available}  \\

\vspace{-0.6cm}
\noindent
{at http://www.cc.saga-u.ac.jp/saga-u/riko/physics/quantum1/structure.html} \\

\vspace{-0.6cm}
\noindent
\normalsize
{or at ftp://ftp.cc.saga-u.ac.jp/pub/paper/riko/quantum1.} \\

\vspace{1.0cm}

\vspace{-0.5cm}
\hfill
{submitted to Australian Journal of Physics}

\vfill\eject
\pagestyle{plain}
\begin{center}
 
\Large
{Shadowing Aspects of Nuclear Parton Distributions} \\
 
\vspace{0.5cm}
 
{S. Kumano $^*$}             \\
 
{Department of Physics}    \\
 
{Saga University}      \\

{Saga 840, Japan} \\

\vspace{0.7cm}

\normalsize
Abstract
\end{center}
\vspace{-0.30cm}

Various shadowing aspects of nuclear parton distributions are
discussed in a parton-model framework.
We explain existing $x$ dependent data of $F_2^A/F_2^D$,
then the model is applied to a flavor asymmetry
$\bar u-\bar d$ in a nucleus,
to shadowing in valence-quark distributions, and to nuclear
dependence of $Q^2$ evolution.
First, we find that a finite nuclear $\bar u-\bar d$ distribution 
could be possible due to nuclear interactions even
in the flavor symmetric case in the nucleon ($\bar u -\bar d$=0).
Second, it is indicated that valence-quark shadowing could be used 
for discriminating among various shadowing models.
Third, we find that nuclear dependence of $Q^2$ evolution 
$\partial [F_2^{Sn}/F_2^C]/ \partial [\ln Q^2]$,
which was found by NMC, is essentially understood by 
modification of parton distributions in nuclei.
However, higher-twist effects in nuclear interactions could
be tested by studying the details of the nuclear $Q^2$ evolution.

\vspace{0.6cm}


\noindent
{\bf 1. Introduction: structure function $\bf F_2$ in nuclei}

\vspace{0.2cm}

Nuclear modification of the structure function $F_2$ has
been investigated extensively.
In recent years, it became possible to measure
the ratio $F_2^A/F_2^D$ in the ``shadowing" region,
namely at small $x$.
Accurate experimental data make it possible to test
the details of theoretical shadowing models.
Among various models for describing $F_2$,
we study a hybrid parton model with $Q^2$-rescaling and 
parton-recombination mechanisms [1].
The rescaling model was originally proposed by considering
possible confinement-radius changes in nuclei.
Although there is still such a possibility 
of the nucleon-size modification,
the major part of the EMC effect
at medium $x$ could be explained by the nuclear binding.
In order to understand the seemingly different ideas,
Close, Roberts, and Ross claim that it is possible
to relate two pictures by using factorization-scale independence [2].
In calculating a nuclear structure function in the operator
product expansion, we separate it into two pieces depending 
on short and long distance physics.
This separation scale, which is called the factorization scale, 
is an arbitrary constant and a physical observable should not depend on it.
The various interpretations correspond to different choices of
the factorization scale.
Therefore, we may view the rescaling explanation as an alternative
way of explaining the EMC effect to the binding explanation
rather than as a different one.

In the small $x$ region, we use the parton-recombination mechanism
for describing the shadowing.
The longitudinal localization size of a parton exceeds
the average nucleon separation in a nucleus at $x<0.1$.
It means that partons in different nucleons could interact 
in the nucleus, and the interaction is called parton recombination.
Its contribution is a higher-twist effect which
is proportional to $\alpha_s/Q^2$. In order to explain 
$x$ dependence of $F_2^{Ca}/F_2^D$ measured by NMC,  
it is necessary to take rather small $Q^2$ ($Q_0^2=$0.8 GeV$^2$)
if the KMRS-B type distributions are chosen as input ones.
The rescaling and recombination mechanisms are used for 
calculating nuclear parton distributions at $Q_0^2$.
Then, they are evolved to those at larger $Q^2$.

We show theoretical results for $x$ dependence
of the ratio $F_2^{Ca}/F_2^D$ [1] together with experimental data.
In Fig. 1, the solid curve shows theoretical results 
at $Q^2$=5 GeV$^2$. They are compared with experimental data
by NMC and SLAC [3]. We obtain good agreement with the data
if $Q_0^2$ is adjusted to a small value (0.8 GeV$^2$). 
The model explains the modification
at small $x$ ($<0.05$) and at large $x$ ($>0.8$) in terms of
the recombinations, the depletion at medium $x$ by the rescaling,
and the antishadowing at $x\approx 0.2$ by competition between
two mechanisms.
We find that our parton model can explain the NMC data $F_2^A/F_2^D$,
so that the model is applied to other interesting topics
in the following sections.

\vspace{0.6cm}

\noindent
{\bf 2. Flavor asymmetric distribution $\bf \bar u-\bar d$ in nuclei}

\vspace{0.2cm}

NMC suggested that the Gottfried sum rule should be violated in 1991.
Since then, there have been efforts to investigate mechanisms of
creating a flavor asymmetric distribution $\bar u-\bar d$ in the nucleon.
In order to test the NMC finding, Drell-Yan experiments for the proton 
and the deuteron are in progress at Fermilab.
On the other hand, there exist Drell-Yan data for various nuclear
targets, so that some people use, for example, tungsten data for 
investigating the flavor asymmetry. 
However, we have to be careful in comparing the NMC results 
with the tungsten data because of possible nuclear medium effects.
In order to find whether such comparison makes sense, 
we estimate a nuclear modification effect. 
It in turn could be found experimentally
by analyzing accurate Drell-Yan data in the near future. 

We investigate the $\bar u-\bar d$ distribution in the tungsten
nucleus [4]. If isospin symmetry could be applied to parton distributions
in the proton and the neutron, the distribution per nucleon becomes
$x[\bar u(x)-\bar d(x)]_A = 
- \varepsilon x [\bar u(x)-\bar d(x)]_{proton}$
without considering nuclear modification. 
It is just the summation of proton and neutron contributions.
The neutron-excess parameter $\varepsilon$ is defined by
$\varepsilon =(N-Z)/(N+Z)$, and 
it is 0.196 for the tungsten $_{74}^{184} W_{110}$.
According to the above equation, the flavor distribution
should be symmetric ($[\bar u-\bar d]_W=0$) if it is symmetric
in the nucleon.
However, it is not the case in the recombination model.
In a neutron-excess nucleus ($\varepsilon >0$) such
as the tungsten, more $\bar d$ quarks are lost than $\bar u$ quarks
are in the parton recombination process 
$\bar q q \longrightarrow G$ in Fig. 2 because of
the $d$ quark excess over $u$ in the nucleus.
In the case of $\bar u-\bar d$=0 in the nucleon,
only $\bar q q \longrightarrow G$ processes contribute.
If it is not symmetric in the nucleon [$(\bar u-\bar d)_N \ne 0$],
$\bar q G\rightarrow \bar q$ type contributions become dominant.
The flavor asymmetry is given by
$x[\Delta \bar u(x) - \Delta \bar d(x)]_A = -(w_{nn} - w_{pp})
x[\Delta \bar u(x) - \Delta \bar d(x)]_{pp}$,
where $w_{nn}$ and $w_{pp}$ are neutron-neutron and proton-proton
combination probabilities ($w_{nn} - w_{pp}=\varepsilon$).
$[\bar u(x)-\bar d(x)]_{pp}$ is the asymmetry produced in the 
proton-proton combination. Detailed equations of the recombination
effects are given in Ref. [4].

We evaluate the recombination contributions with the input
parton distributions MRS-D0 (1993) at $Q^2$=4 GeV$^2$.
In the $(\bar u-\bar d)_N = 0$ case, 
$\Delta=0$ is taken in the MRS-D0 distribution.
Obtained results are shown in Fig. 3, where the solid (dashed) 
curve shows the $x[\Delta\bar u-\Delta\bar d]_A$ distribution 
of the tungsten nucleus with the flavor symmetric (asymmetric)
sea in the nucleon.
In the $(\bar u-\bar d)_N = 0$ case, the positive contribution
at small $x$ can be understood by the processes in Fig. 3.
In the $(\bar u-\bar d)_N \ne 0$ case, 
the $\bar q(x) G\rightarrow \bar q$ process
is the dominant one kinematically at small $x$.
Its contribution to $\bar q(x)$ is negative because $\bar q$
with momentum fraction $x$ is lost in the recombination process.
However, more $\bar d(x)$ is lost than $\bar u(x)$ is because
of the $\bar d$ excess over $\bar u$ in the proton
($[\Delta\bar u-\Delta\bar d]_{pp}>0$).
Therefore, the overall contribution becomes negative
in $[\Delta \bar u(x) - \Delta \bar d(x)]_A$
due to the neutron excess as shown in Fig. 3.
In the medium $x$ region, the $\bar q G\rightarrow \bar q(x)$
process becomes kinematically favorable. Because it produces
$\bar q$ with momentum fraction $x$, its contribution becomes
opposite to the one at small $x$.
The above results are obtained at $Q^2$=4 GeV$^2$. 
Considering the factor of two coming from the $Q^2$ dependence,
we find that the nuclear modification is of the order of 2\%--10\%
compared with the asymmetry suggested by the MRS-D0 distribution.

{\bf [Summary]}
We find that finite flavor asymmetric distributions 
are possible in nuclei ($[\bar u(x)-\bar d(x)]_A \ne 0$)
even in the symmetric case in the nucleon ($[\bar u(x)-\bar d(x)]_N=0$).
According to the recombination model, the nuclear effects
on $[\bar u(x)-\bar d(x)]_A$ are of the order of 2\%--10\% compared with
the one estimated by the NMC flavor asymmetry in the nucleon.
Because the Drell-Yan experiments on proton, deuteron, and nuclear targets
will be completed at Fermilab in the near future, 
it is possible in principle
to study the nuclear modification experimentally.

\vspace{0.6cm}

\noindent
{\bf 3. Valence-quark shadowing}

\vspace{0.2cm}

It is shown in section 1 that the parton-recombination model
explains the $F_2$ shadowing fairly well if small $Q_0^2$ is taken.
However, it is not the only shadowing explanation.
There exist other models which produce similar results 
on the $F_2$ shadowing.
We may discriminate among various models by predicting unobserved
quantities. Gluon shadowing [5] should be an interesting one for
future theoretical and experimental studies. 
As another possible quantity for testing the models,
we discuss valence-quark shadowing in this section [6].
We show that it depends much on theoretical ideas
by using two different shadowing models.
The first one is the recombination model with $Q^2$ rescaling
in section 1. The second is an aligned-jet model [7], which is 
an extension of the vector-meson-dominance (VMD) model.

The first model prediction is the following.
The rescaling produces depletion in the ratio 
$R_3\equiv V_A/V_D$ at medium $x$ as it explains
the original EMC effect. Because the rescaling satisfies
the baryon-number conservation, the ratio $R_3$ becomes larger
than unity at small $x$. 
Parton-recombination contributions are rather contrary to
those of the rescaling. The recombinations decrease the ratio 
at small $x$ and increase it at medium and large $x$.
The overall contributions are shown in Fig. 4 by a solid
curve (model 1). 
Antishadowing is obtained in this model instead of
the shadowing in $F_2$. This difference is caused by the fact that
the rescaling is used only for the valence-quark distribution and
not for sea-quark and gluon distributions.

In the aligned-jet model, the virtual photon transforms into 
a $q\bar q$ pair, which then interacts with the target.
However, the only $q\bar q$ pair aligned in the direction 
of $\gamma$ interacts in a similar way
to the vector-meson interaction with the target.
The propagation length of the hadronic ($q\bar q$) fluctuation
exceeds the average nucleon separation in a nucleus
at small $x$, so that the shadowing phenomena occur
due to multiple scatterings.
In this model, vector-meson-like $q\bar q$ pairs
interact with sea quarks and valence quarks in the same manner. 
Therefore, the valence-quark shadowing is very similar to 
the $F_2$ one as shown in Fig. 4 by a solid curve (model 2) [7].
The shadowing is calculated by the aligned-jet model at small $x$,
then the curve is extrapolated into the medium $x$ region simply
by using the baryon-number conservation.
It is interesting to find completely different
shadowing results in models 1 and 2.

Next, we investigate whether both results are allowed
by existing experimental data.
There are neutrino data for the deuteron
and for nuclei; however, the accuracy is not good enough
to test the nuclear modification particularly in the
small $x$ region. Instead, we use existing $F_2^A/F_2^D$ data
and the baryon-number conservation.
We assume that $V_A/V_D$ is the same as $F_2^A/F_2^D$ in
the region $x \ge 0.3$ because $F_2$ is dominated by
valence-quark distributions.
The SLAC $F_2^{Ca}/F_2^D$ data [3] at $x>0.3$ are fitted 
by a smooth curve in Fig. 4,
then it is extrapolated into the small $x$ region by considering
the baryon-number conservation.
Because there is no guideline for the extrapolation,
a straight dashed-line is simply drawn in Fig. 4.
As another possibility, the curve is smoothly extrapolated 
by allowing about 6\% antishadowing at $x=0.1-0.2$ (dotted curve). 
The hatched area between the line and the curve
is roughly the region, which is allowed
by the current experimental data.
The figure indicates that both results are allowed 
by the experimental data at this stage.

{\bf [Summary]}
The valence-quark shadowing is investigated in two different models: 
the parton-recombination with $Q^2$ rescaling and the aligned-jet model.
Both shadowing results are completely different, so that they can 
be tested by accurate experimental measurements.
However, both theoretical results are allowed by 
the existing experimental data at this stage.
The shadowing could be studied at HERA in future by observing 
charged pion productions [8]; hence, there is a good possibility
of discriminating among various shadowing models by
studying the valence-quark distributions.

\vspace{0.6cm}

\noindent
{\bf 4. Nuclear dependence of $Q^2$ evolution}

\vspace{0.2cm}

The $x$ dependence of nuclear modification in the structure function
$F_2$ has been accurately measured and it is well studied
theoretically. On the contrary, there is little investigation on
$Q^2$ dependence partly because experimental accuracy was not
good enough to find nuclear dependence of $Q^2$ evolution.
However, it becomes possible to find the details of
nuclear $Q^2$ evolution due to recent NMC analysis 
of tin and carbon $F_2$ ratios [9].
They found significant deviations 
$\partial [F_2^{Sn}/F_2^C]/ \partial [\ln Q^2] \ne 0$,
so that it is an interesting topic for theoretical studies.

$Q^2$ dependence of structure functions can be calculated
by using the Dokshitzer-Gribov-Lipatov-Altarelli-Parisi (DGLAP) equations.
In addition, parton-recombination (PR) contributions are expected
at small $x$. 
Although the DGLAP equations are well tested by 
various experimental data, the PR equations 
are not well established yet.
There are two possible sources for the nuclear dependence in 
the evolution equations.
One is nuclear parton distributions, 
and the other is the recombination effects.
As it is discussed in sections 1, 2, and 3, parton
distributions are modified due to nuclear medium effects.
The modifications affect the $Q^2$ evolution through
splitting functions.
Because parton-recombination probability is proportional to
$A^{1/3}$, which is the number of nucleons in the longitudinal 
direction, the recombination effects 
could become significant in large nuclei. 
We study these two nuclear contributions [10]
in comparison with the NMC data for
$\partial [F_2^{Sn}/F_2^C]/ \partial [\ln Q^2]$.

As the input parton distributions in tin and carbon nuclei, 
we employ those in section 1 [1]. 
They are evolved by leading-order (LO) DGLAP,
next-to-leading-order (NLO) DGLAP, and PR
equations with the help of the computer program bf1.fort77 
in Ref. [11]. 
Three evolution results are shown at $Q^2$=5 GeV$^2$ 
together with the preliminary NMC data [9]
in Fig. 5.
The DGLAP results agree with the experimental tendency; however,
the PR curve is well below the data points at small $x$.
The disagreement between the PR results and the NMC data
does not mean the PR mechanism is in danger.
The large discrepancy is caused first by the choice of
initial $Q^2$ ($Q_0^2$=0.8 GeV$^2$)
and second by the choice of the constant $K_{HT}$=1.68 in
a higher-dimensional gluon distribution. 
Although the value of $K_{HT}$ is taken according to 
the Qiu's numerical analysis, it is not a uniquely determined quantity. 
In order to discuss the validity
of the parton recombination effects, we have to estimate
$K_{HT}$ theoretically. 
From Fig. 5, we can at least rule out large recombination
contributions. 
Because the figure indicates significant recombination effects,
the $Q^2$ derivative could be used for examining such contributions.

{\bf [Summary]}
The NMC's nuclear $Q^2$ dependence, 
$\partial [F_2^{Sn}/F_2^C]/ \partial [\ln Q^2]\ne 0$,
could be essentially understood by ordinary $Q^2$ evolution
equations together with modified nuclear parton distributions.
Because the PR evolution results disagree with the data,
{\it large} higher-twist effects from the parton recombinations 
could be ruled out.
However, it is encouraging to study the details of the
recombination mechanism in comparison with the NMC data.

\vspace{0.6cm}
\noindent
P.S. 
The Adelaide group for hadron structure 
(R. P. Bickerstaff, W. Melnitchouk,
     K. Saito, A. W. Schreiber, A. I. Signal, and A. W. Thomas)  
has made many contributions to similar topics
on the nuclear structure functions
and on the flavor asymmetry.


\begin{center}
{\bf Acknowledgments} \\
\end{center}
\vspace{-0.2cm}

S. K. thanks the Institute of Theoretical Physics
at University of Adelaide for its financial support 
for his participating in this conference.
This research was partly supported by the Grant-in-Aid for
Scientific Research from the Japanese Ministry of Education,
Science, and Culture under the contract number 06640406.

$~~~$

\noindent
{* Email: kumanos@cc.saga-u.ac.jp. 
   Information on his research is available}  \\

\vspace{-0.6cm}
\noindent
{at http://www.cc.saga-u.ac.jp/saga-u/riko/physics/quantum1/structure.html.} \\

\vspace{0.1cm}

\begin{center}
{\bf References} \\
\end{center}

\vspace{0.3cm}
 
\vspace{-0.30cm}
\vspace{-0.38cm}
\begin{description}{\leftmargin 0.0cm}

\vspace{-0.38cm}
\item{[1]}
S. Kumano, Phys. Rev. C48 (1993) 2016; C50 (1994) 1247.

\vspace{-0.38cm}
\item{[2]}
F. E. Close, R. G. Roberts, and G. G. Ross,
           Nucl. Phys. B296 (1988) 582.

\vspace{-0.38cm}
\item{[3]}   
P. Amaudruz et al. (NMC collaboration), Nucl. Phys. B441 (1995) 3;
J. Gomez et al. (SLAC-E139 collaboration), Phys. Rev. D49 (1994) 4348.

\vspace{-0.38cm}
\item{[4]}
S. Kumano, Phys. Lett. B342 (1995) 339.

\vspace{-0.38cm}
\item{[5]}
S. Kumano, Phys. Lett. B298 (1993) 171.

\vspace{-0.38cm}
\item{[6]}
R. Kobayashi, S. Kumano, and M. Miyama, 
              Phys. Lett. B354 (1995) 465. 

\vspace{-0.38cm}
\item{[7]}
L. L. Frankfurt, M. I. Strikman, and S. Liuti,
                  Phys. Rev. Lett. 65 (1990) 1725.

\vspace{-0.38cm}
\item{[8]}
S. Kumano, nucl-th/9510029.

\vspace{-0.38cm}
\item{[9]}
A. M\"ucklich and A. Sandacz (NMC), private communications on the
NMC preliminary data for $\partial [F_2^{Sn}/F_2^C]/ \partial [\ln Q^2]$.

\vspace{-0.38cm}
\item{[10]}
S. Kumano and M. Miyama, hep-ph/9512244.

\vspace{-0.38cm}
\item{[11]}
M. Miyama and S. Kumano, hep-ph/9508246, Comput. Phys. Commun. in press;
R. Kobayashi, M. Konuma, and S. Kumano, 
              Comput. Phys. Commun. 86 (1995) 264.

\end{description}

\vfill\eject

$~~~$

\vspace{-6.0cm}
\hspace{+0.5cm}
\epsfxsize=12.0cm
\epsfbox{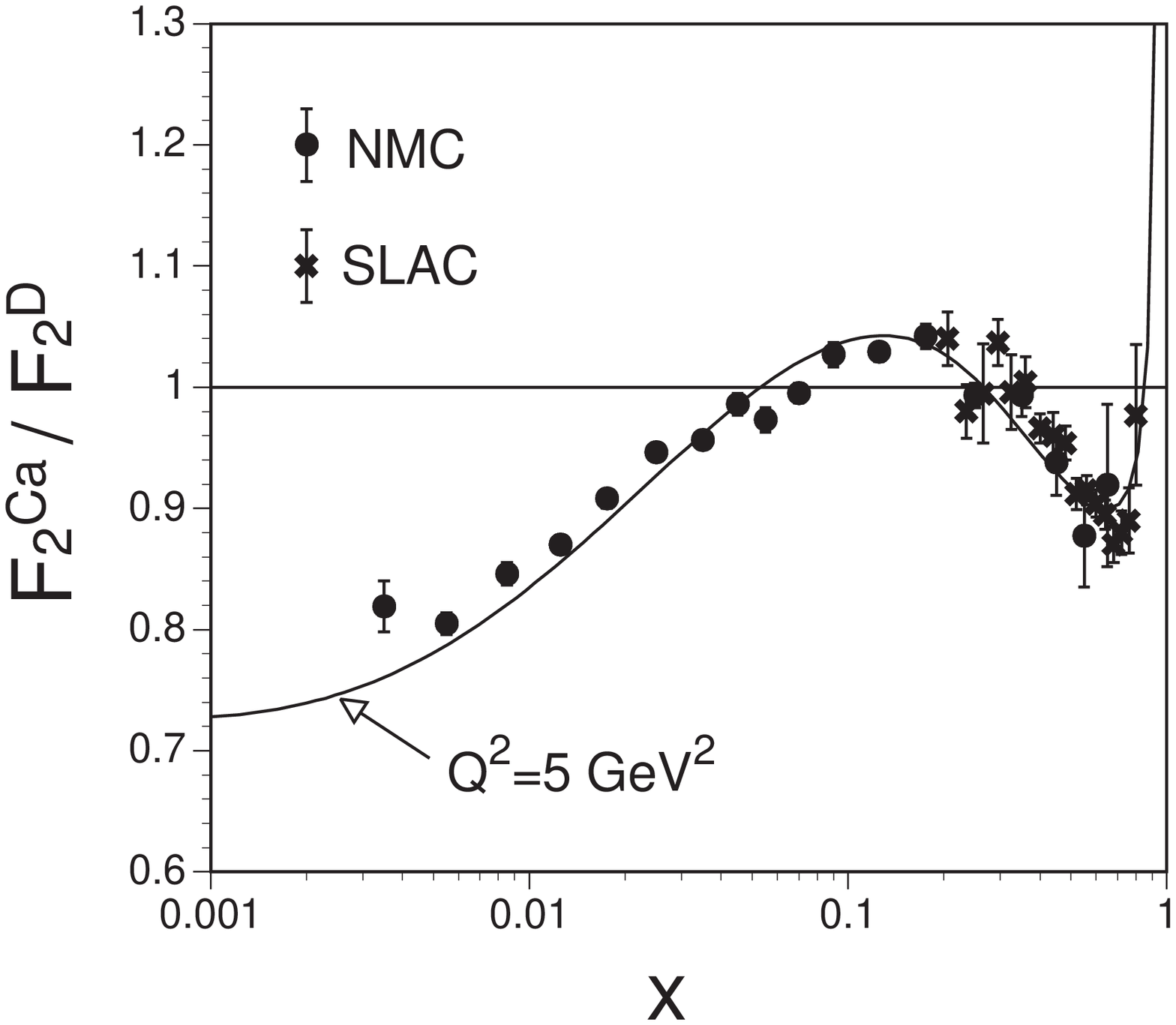}

\vspace{-2.0cm}
\centering{\large{Fig. 1 \ $x$ dependence of $F_2^{Ca}/F_2^D$.}}

\vspace{+4.0cm}
\hspace{+0.0cm}
\epsfxsize=12.0cm
\epsfbox{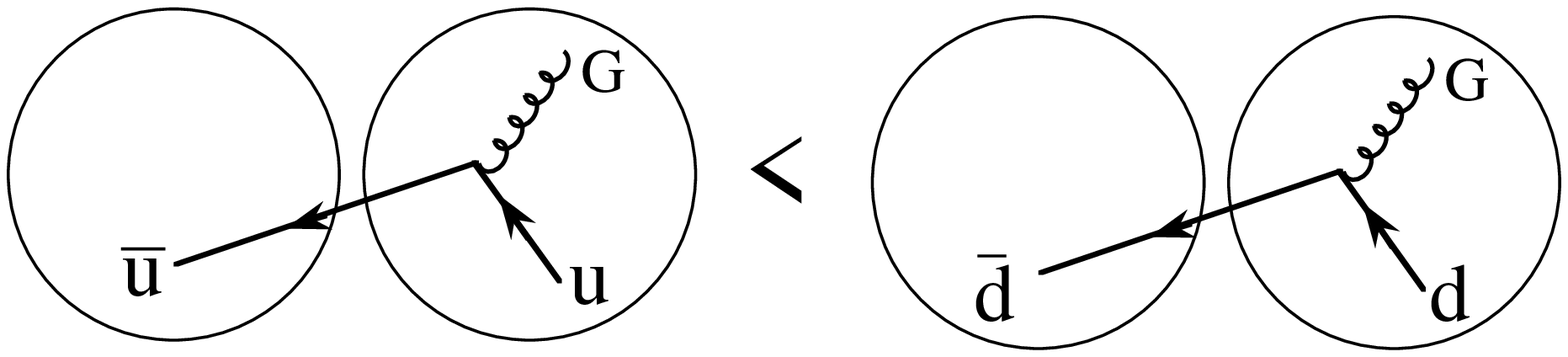}

\vspace{+2.5cm}
\centering{\large{Fig. 2 \ Parton recombinations.}}

\vfill\eject

$~~~$

\vspace{-3.0cm}
\hspace{-0.5cm}
\epsfxsize=10.5cm
\epsfbox{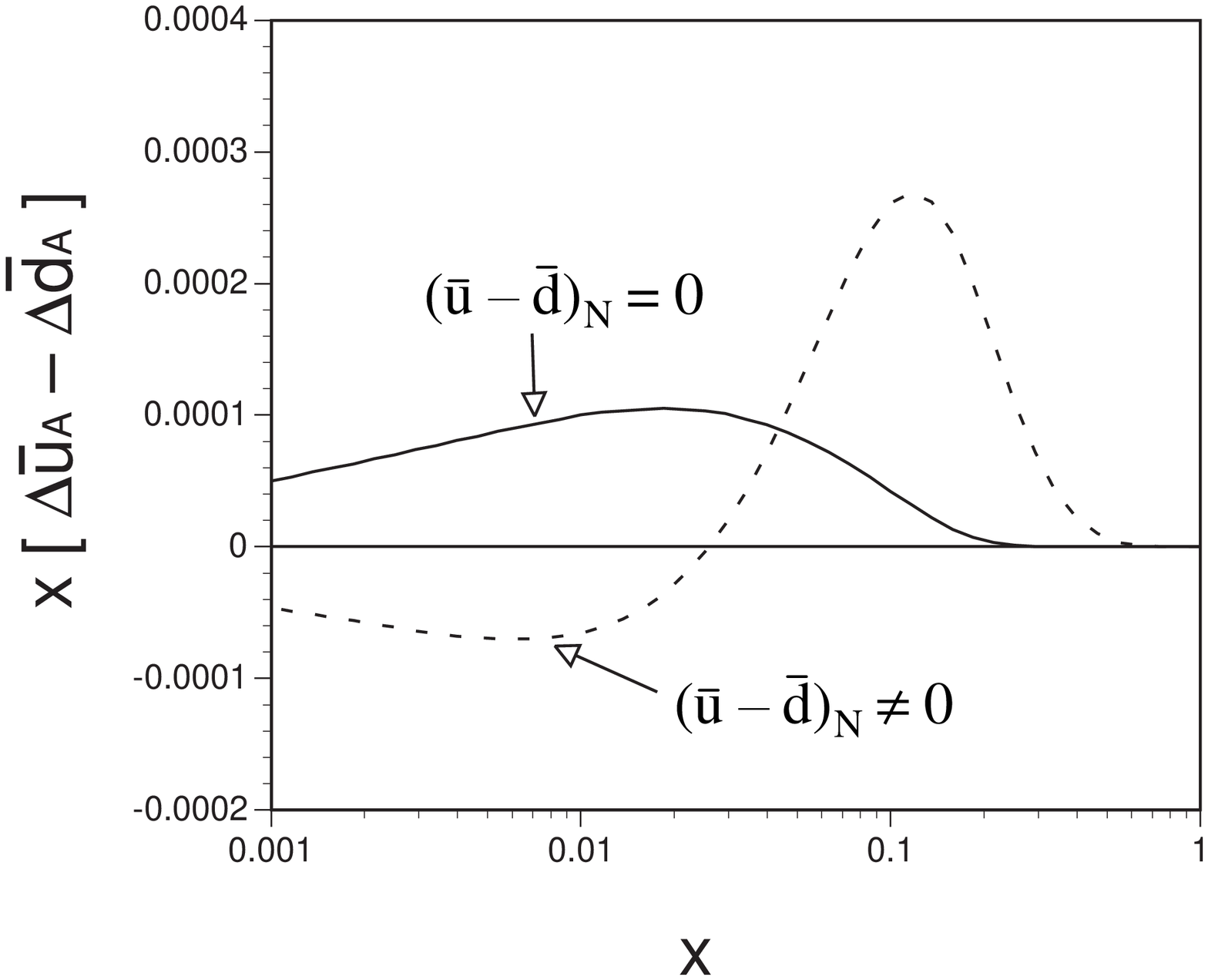}

\vspace{-5.0cm}

\vspace{-0.5cm}
\hspace{-0.5cm}
\epsfxsize=10.0cm
\epsfbox{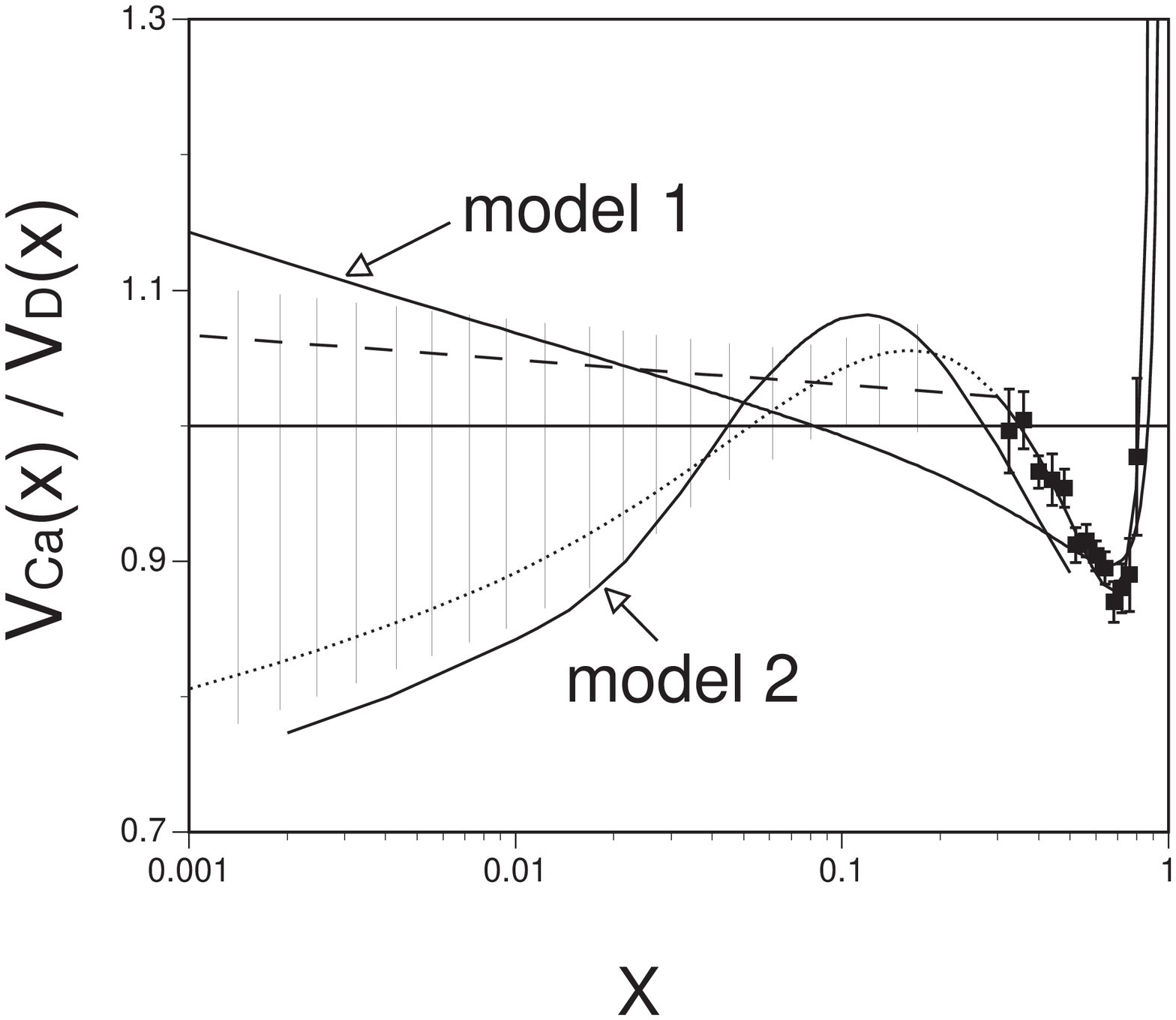}

\vspace{-1.0cm}
\centering{\large{Fig. 4 \ Valence-quark ``shadowing".}}

\vspace{-13.5cm}
\centering{\large{Fig. 3 \ Flavor asymmetry in W.}}

\vfill\eject

$~~~$

\vspace{-2.0cm}
\hspace{-0.5cm}
\epsfxsize=10.5cm
\epsfbox{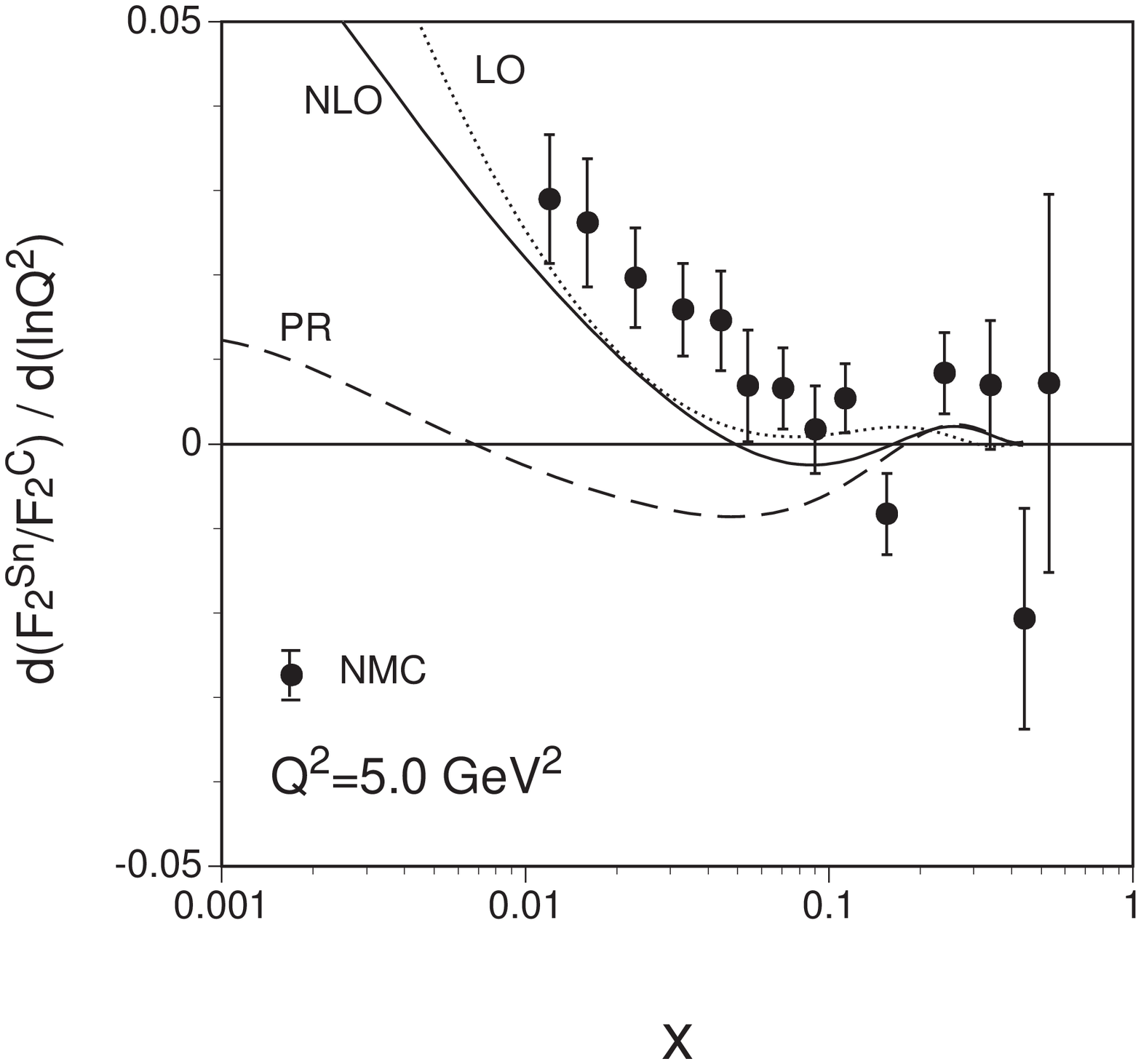}

\vspace{-3.5cm}
\centering{\large{Fig. 5 \ Nuclear $Q^2$ evolution.}}

\end{document}